\documentclass[12pt]{iopart}

\usepackage{graphics,graphicx}
\usepackage{harvard}
\def\apj{\rm ApJ}
\def\apjl{\rm ApJL}

\def\aj{\rm AJ}
\def\mnras{\rm MNRAS}
\def\nat{\rm Nature}

\def\aap{\rm AAP}
\def\araa{\rm ARA\&A}
\newcommand{\msun}{M$_{\odot}$}
\def\beq{\begin{equation}}
\def\eeq{\end{equation}}

\begin{document}

\title{Black Holes in the Early Universe}

\author{Marta Volonteri$^{1,2}$, Jillian Bellovary$^2$}

\address{$^1$Institut d'Astrophysique de Paris, Paris, France\\ $^2$University of Michigan, Ann Arbor, MI, USA}
\ead{martav@iap.fr; jillianx@umich.edu}

\begin{abstract}

\noindent
The existence of massive black holes was postulated in the sixties,
when the first quasars were discovered. In the late nineties their
reality was proven beyond doubt, in the Milky way and a handful nearby
galaxies. Since then, enormous theoretical and observational efforts
have been made to understand the astrophysics of massive black
holes. We have discovered that some of the most massive black holes
known, weighing billions of solar masses, powered luminous quasars
within the first billion years of the Universe. The first massive
black holes must therefore have formed around the time the first stars
and galaxies formed. Dynamical evidence also indicates that black
holes with masses of millions to billions of solar masses ordinarily
dwell in the centers of today's galaxies. Massive black holes populate
galaxy centers today, and shone as quasars in the past; the quiescent
black holes that we detect now in nearby bulges are the dormant
remnants of this fiery past.  In this review we report on basic, but
critical, questions regarding the cosmological significance of massive
black holes. What physical mechanisms lead to the formation of the
first massive black holes? How massive were the initial massive black
hole seeds? When and where did they form? How is the growth of black
holes linked to that of their host galaxy? Answers to most of these
questions are work in progress, in the spirit of these Reports on
Progress in Physics.
\end{abstract}

\section{Overview: Evidences and Constraints on Black Holes in the First Galaxies}

Since the discovery of Sagittarius A$^*$ \cite{Balick74}, evidence has
been growing that all massive galaxies host massive black holes (MBHs)
in their centers.  As our observing power has increased, we have seen
how these MBHs can exist as powerful quasars, capable of discharging
prodigious amounts of energy.  Over time, we have formed a picture
where MBHs and their host galaxies co-exist and co-evolve.  Galaxies
build up their mass through accretion and mergers, and during these
events MBHs build up their mass as well, undergoing periods of
activity and quiescence \cite{Croton06,Bower06}.  An apparent result
of this hierarchical buildup is the observed relation between the host
galaxy spheroid and the MBH mass.  This relation holds whether one
measures the spheroid's velocity dispersion\footnote{The velocity
dispersion is the root mean square of stellar velocities, and it gives
a measure of the motion of stars under the effect of the galaxy
potential.} \cite{Ferrarese00,Gebhardt00,Tremaine02,Gultekin09}, mass
\cite{Magorrian98,Haring04}, luminosity \cite{Marconi03}, or other
host properties (such as number of globular clusters \cite{Burkert10}
and their global velocity dispersion \cite{Sadoun2012} or the Sersic
index \cite{Graham2007}).

To determine how these relationships (and MBH growth and evolution in
general) have arisen, we must look to the early universe.  The
discovery of quasars at $z \sim 6$ demonstrates that MBHs must form
extremely early on and grow rapidly in order to acquire $10^9$ \msun~
of mass within a span of $\sim 1$ Gyr \cite{Fan01}.  How effectively
MBH growth proceeds from their `seeds' is hard to gauge, as most
current observations are sensitive only to the most luminous quasars
($L> 10^{44}$ erg/s), hence the most massive MBHs ($M_{BH}>
10^{8-9}$\msun). Recent attempts at investigating lower luminosity
sources have provided conflicting results. While \cite{Treister11}
suggested that the vast majority of high redshift MBHs may be obscured
and accreting rapidly, hinting that much of MBH growth may be hidden
from view, \cite{Willott2011}, \cite{Cowie2012} and \cite{Fiore2012}
do not find a population of obscured low-luminosity Active Galactic
Nuclei (AGN).

As we expand our depth and wavelength coverage of the universe, we can
use the evolving luminosity functions of quasars and galaxies to
constrain MBH activity, and the link between MBHs and their hosts.
Several groups have found evidence for ``downsizing,'' in which the
peak of quasar activity was dominated by brighter, more massive MBHs
at higher redshifts \cite{Croom09,Willott10,Assef11,Glikman11}.  Such
a pattern also reflects that of star formation activity
\cite{Cowie96,Kodama04,Treu05,Bundy06}, and strengthens the argument
that MBH activity and star formation/galaxy growth are interconnected.





 In section 2 we will discuss the potential physical mechanisms which
 may form MBHs, and how future observations may aid us in determining
 the actual formation pathway.  In section 3 we describe the numerous
 ways in which MBHs are interconnected with their host galaxies.
 Section 4 chronicles the current status of cosmological simulations,
 which are a crucial tool for exploring the entire co-evolutionary
 history of MBHs and their hosts.  Finally, in section 5 we summarize
 with some open questions in the field and the future outlook of each.

\section{The Physics of Black Hole Formation}

As discussed in the previous section, MBHs with masses of $10^9$
\msun~ have been identified at distances corresponding to a light
travel time of more than 12 billion years. This means that these MBHs,
as massive as the most massive black holes we detect in today's
galaxies, were in place when the Universe was less than one billion
years old.  One can easily understand why this is a significant feat
by estimating the growth time of an MBH. Typically it is assumed that
MBHs cannot accrete at rates much higher than permitted by the
``Eddington limit", which occurs when the luminosity of a source
becomes large enough that radiation pressure overcomes gravity. Above
the Eddington limit, material is pushed away instead of falling
towards the black hole to fuel activity.  The Eddington luminosity can
be written as $L_{\rm Edd}=M_{BH}c^2/t_{\rm Edd}$, where $t_{\rm
Edd}=\frac{\sigma_T \,c}{4\pi \,G\,m_p}= 0.45$ Gyr (here $c$ is the
speed of light, $\sigma_T$ is the Thomson cross section, and $m_p$ is
the proton mass). Therefore, if the inflow rate of mass towards the
MBH is $\dot{M}_{in}$, and $\dot{M}$ is the mass that goes into
increasing the MBH mass, from: \beq L=\epsilon \, \dot{M}_{in} c^2=
L_{\rm Edd} c^2 \eeq and $dM = (1-\epsilon) dM_{in}$, where
$\epsilon\simeq 0.1$ is the efficiency of conversion of rest-mass into
energy.  The growth time of an MBH with initial mass $M_0$ is
 
\beq
t_{\rm growth}=t_{\rm Edd} \frac{\epsilon}{1-\epsilon}\ln \left(\frac{M_{BH}}{M_0}\right).
\eeq

For MBHs to reach $10^9$ \msun~ within 1 Gyr, they must form early on
and grow very rapidly \cite{Haiman01}.  To accomplish this, it is
helpful if they form with an ``intermediate'' mass -- $M_0$ between
100 -- 10$^5$ \msun~ or so.  A black hole of 1 \msun~ may grow 9
orders of magnitude in slightly over 1 Gyr only if it accretes at the
Eddington rate for its entire lifetime; if feedback effects from stars
and the MBH are negligible this growth may be possible, but such a
scenario is extremely unlikely.

The key to forming an MBH seed at high redshift is obtaining the rapid
collapse of `baryons'. Stars and gas represent the ÔbaryonicÕ
content of galaxies, to be contrasted with non-baryonic dark matter
that does not interact radiatively, but only gravitationally with its
environment. In the standard picture, the mass content of the Universe
is dominated by cold dark matter, with baryons contributing at a 10\%
level only. Starting from a Gaussian density fluctuation field in a
quasi-homogeneous Universe, dark matter perturbations grow to the
point they collapse and virialize forming self gravitating halos
within which gas eventually condenses to form the luminous portion of
galaxies, comprising both gas and stars formed out of this
gas. Normally, during the formation of galaxies part of the gas cools
and becomes dense enough that gas clouds ``fragment" into clumps where
stars eventually form.

In the event of a rapid collapse of gas, which represents one of the
possible starting points for MBH formation, some mechanism must be in
place to prevent fragmentation into small clumps (and subsequent star
formation).  If fragmentation is prevented, most of the gas is
actually funneled to forming the black hole, and the resulting object
is massive.  If the baryons are predominantly stellar objects, they
must be able to merge in an efficient way without ejecting significant
mass from the system.  In all cases, the effects of feedback from
accretion onto the MBH or nearby star formation will surely play a
role in how massive the MBH can become before its growth is limited by
its environment.  In the following sections we review several proposed
formation scenarios for MBH seeds and discuss the physical
implications of each.

\subsection{Population III Stars}

For several years, the most promising mechanism to form
MBHs at high redshift was assumed to be via the remnants of the very first generation of stars
\cite{Madau01}. The first bout of star formation must occur, by
definition, when gas still had primordial composition (i.e., all the
atoms heavier than hydrogen were produced through primordial
nucleosynthesis, and heavy elements, globally defined as ``metals" in
astrophysics, were absent). These stars are commonly referred to as
Population III, or Pop III, stars.  Such stars have been speculated to
have formed with masses much larger than today's stars, so that the
typical mass of a Pop III star would be hundreds of solar masses
\cite{Couchman86,Abel02,Bromm04}. In particular, if stars of
primordial composition existed with masses greater than 260 \msun,
they are predicted to directly collapse into a black hole of $\sim
100$ \msun~ \cite{Bond84,Heger02}.

Recent simulation results have put this picture into question.  The
inclusion of more complex physics combined with higher resolution has
resulted in models of Pop III stars with lower masses.  \cite{Turk09}
showed that fragmentation may be common in Pop III star collapse, and
thus Pop III binaries with masses of a few tens of \msun~ could be
fairly run-of-the-mill.  Improvements in simulation techniques such as
the addition of turbulence \cite{Clark11}, radiative feedback
\cite{Stacy12}, or entirely new codes \cite{Greif11} have confirmed
this result, showing that Pop III stars may form in binaries and/or
clusters with a wide range of initial masses (though primarily in the
10-100 \msun~ range, probably).  The existence of a large number of
isolated stars with masses larger than 260 \msun~ has thus been called
into doubt.  While the remnants of stars with masses less than 140
\msun~ might still become MBH seeds, these objects form in shallow
potential wells which are unable to retain photoionized heated
gas. and may be starved of accretable gas for a local
Hubble time \cite{Johnson07}.  Massive Population III binaries may be
able to fuel an existing MBH via binary accretion/ejection and
subsequent tidal disruption (see \cite{Bromley12}), but the central stellar density 
may not be large enough for this process to be
efficient until a large number of stars have formed (in simple models the event rate 
scales linearly with the central stellar density, but see \cite{Merritt2009} and references 
therein for detailed calculations). Models which predict the evolution of Pop III seeds into
today's MBH population (i.e. \cite{Volonteri03}) assume that a
substantial fraction of Pop III stars form MBH seeds.  The potential
rarity of Pop III MBH seeds combined with their smaller size makes
them rather unattractive candidates for being the precursors to $z =
6$ quasars.


\subsection{Direct Collapse and Quasi-Stars}

Several works have recently discussed how a rapid inflow of gas may
collapse to form an MBH directly
\cite{Loeb94,Eisenstein95,Oh02,BrommLoeb2003,Koushiappas04,Begelman06,Lodato06,Begelman08}.
Such a phenomenon can only occur if the gas does not fragment, but
instead undergoes larger-scale dynamical instabilities.  Preventing
fragmentation generally requires inhibiting cooling, either by
requiring no metals and/or preventing the formation of H$_2$ via a
substantial Lyman-Werner background.  In addition, the gas must have
low intrinsic angular momentum, so that it may reach the halo center
without being inhibited by the formation of a rotationally supported
disk.  In low-angular momentum halos, the disk will be compact, and
gravitational instabilities are able to occur.  These instabilities
are responsible for rapidly funneling gas inwards (and angular
momentum outwards).

Global dynamical instabilities, such as the ``bars within bars''
instability, may act to repeatedly transport gas inwards on the order
of a dynamical time \cite{Begelman06}.  Alternatively, large-scale
torques from major merger activity may be the driver of rapid gas
inflow \cite{Mayer10}.  In such a case, turbulence may be the
inhibitor of fragmentation, and the requirement of metal-free gas may
be relaxed (see also \cite{Begelman09}).

Local dynamical instabilities are instead described through the Toomre
criterion:
\begin{equation}
Q_c = \sqrt{2} \frac{c_s V_h}{\pi G \Sigma R}
\end{equation}

where $c_s$ is the sound speed of the gas, $V_h$ is the circular
velocity of the disk, $\Sigma$ is the surface density of the disk, and
$R$ is the cylindrical radial coordinate.  When $Q_c$ approaches a
value of an order unity, the disk becomes unstable and fragments.
This process is expected to occur in halos with a virial temperature
of $T_{vir} \sim 10^4$ K, which are likely metal-free and do not
contain molecular hydrogen.  In such halos, fragmentation is
suppressed, cooling proceeds gradually, and the gaseous component can
cool and participate in MBH formation before it is turned into stars
\cite{Lodato06}. These halos may need to exist in regions of
ultracritical UV radiation in order to form MBHs by direct collapse
\cite{Dijkstra08,Shang10}, since the average estimated UV background
may not be sufficient to prevent some Pop III stars from forming in
halos of this size \cite{Johnson08}.

In general the step between ``gas infall" and black hole formation has
been dodged, and generic ``post-Newtonian instabilities" have been
advocated for converting the infalling gas into an MBH seed. An
exception is the picture proposed by \cite{Begelman06}, which can
emerge in case of very high gas infall rates, exceeding about 1 \msun~
yr$^{-1}$. In this case, the collapsing gas traps its own radiation
and forms a quasistatic, radiation pressure-supported supermassive
star, which burns hydrogen for about a million years while growing to
a mass $\sim 10^6$ \msun~ \cite{Begelman06,Begelman2010} before its
core collapses and forms a black hole. This black hole is surrounded
by the massive envelope created by the inflow, and it can grow rapidly
at a rate set by the Eddington limit for the massive gaseous envelope,
thus accreting at several hundred to thousand times its own Eddington
rate without violating the luminosity criterion. The resulting object,
comprising a growing MBH embedded in a radiation-pressure supported
envelope, dubbed a `quasistar', resembles a red giant star with a
luminosity comparable to an AGN. As the black hole grows inside it,
its photosphere expands and cools until it hits a minimum temperature
associated with the Hayashi track, at which point it disperses,
leaving behind the naked seed MBH with a mass of thousands to hundred
thousands of solar masses.

\subsection{Collapsing Clusters}

Stellar-dynamical processes may also have a role in creating MBH seeds
at high redshift.  In many cases, galaxies which host MBHs also host
nuclear star clusters \cite{Seth08,Graham09}.  While the link between
these nuclear objects is not fully understood, one can easily imagine
a scenario where a central cluster undergoes core collapse and forms a
single massive compact object at its center.

\cite{Devecchi09} postulate that such a scenario is likely if a
nuclear cluster forms out of the second generation of stars in a
galaxy, when the metallicity (i.e., the heavy element content) of the
gas is still very low, of order $10^{-5}-10^{-4}$ the metallicity of
the Sun (see also \cite{Omukai08}). At sub-solar metallicity, the mass
loss due to winds is much more reduced in very massive stars, which
greatly helps in increasing the mass of the final remnant. In
Toomre-unstable proto-galactic discs, instabilities lead to mass
infall instead of fragmentation into bound clumps and global star
formation in the entire disk. The gas inflow increases the central
density, and within a certain compact region star formation ensues and
a dense star cluster is formed. At metallicities $\sim
10^{-5}-10^{-4}$ solar, the typical star cluster masses are of order
$10^5$ \msun~ and the typical half mass radii $\sim 1$ pc\footnote{A
parsec is a unit of measure of distance corresponding to a parallax of
one second.  It corresponds to 3.26 light-years or $3.1\times 10^{18}$
cm.} \cite{Devecchi2010,Devecchi2012}.  Most star clusters can go into
core collapse in $\sim$ 3 Myr, and runaway collisions of stars form a
very massive star, leading to an MBH remnant with mass $\sim 10^3$
\msun.

A scenario which invokes no dependence on metallicity or redshift has
been proposed by \cite{Davies11}.  If the rapidly inflowing gas
scenario (as mentioned above) occurs in a system where there is a
pre-existing dynamically segregated nuclear cluster, the potential
well of the cluster will deepen rapidly.  In such a situation, the
timescale for collapse of the the compact objects at the core of the
cluster will be shorter than the timescale for dynamical heating via
binaries.  The binding energy of the cluster will be large enough that
recoiling or dynamically ejected objects are actually retained.  The
result is a runaway collapse of compact cluster objects which may form
an MBH of $10^5$ \msun~ at high redshift.

\subsection{Other Models}
Primordial black holes may also be formed in the early universe before
the epoch of galaxy formation \cite{Carr2003,Khlopov2010} in regions
where high density fluctuations are large and the whole region can
collapse and form a primordial black hole.  Generically, primordial
black holes are formed with masses that roughly equal the mass within
the particle horizon at the redshift of their formation
\cite{Zeldovich1967,Hawking1971}. The masses of primordial black holes
therefore range roughly from the Planck Mass (black holes formed at
the Planck epoch) to $\simeq$ \msun~ (black holes formed at the QCD
phase transition) up to $10^5$ \msun~ \cite{Khlopov}.

Several physical or astrophysical constraints restrict the mass range
where primordial black holes are allowed. Primordial black holes with
an initial mass smaller than about $5 \times 10^{14}$~g are expected
to have already evaporated due to Hawking radiation. For masses $\sim
10^{15}$~g, there are strong bounds from the observed intensity of the
diffuse gamma ray background \cite{Page1976}, limiting their
contribution to the matter density to less than one part in $10^8$.
For larger masses, constraints can be deduced from microlensing
techniques \cite{Alcock2000,Tisserand2007} and from spectral
distortions of the cosmic microwave background \cite{Ricotti2008}
which limit the mass to below $\sim 10^3$ \msun.

\subsection{Observational perspectives.}

Directly observing the MBH formation process must wait for the launch
of low-frequency gravitational wave detectors (e.g., eLISA, ET-
Einstein Telescope) or until the next generation of telescopes (or
perhaps the generation after that).  To determine which (if any) of
the seed models is correct, detailed population modeling is required.
The primary parameters which characterize each seed model are
formation redshift, initial mass, and the efficiency of formation.  In
addition, one must consider subsequent effects that affect an MBH's
evolution, such as the growth for extended periods of time that
affects MBHs over the cosmic history, thus erasing information on the
original seed mass, or how central MBH populations are affected by
ejections due to gravitational recoil during MBH mergers, thus erasing
information on the efficiency of MBH formation.  The list of unknowns
is large, but we can explore each of them in turn and hope to
constrain MBH populations vis a vis various seed models.

\begin{figure}[htb]
\includegraphics[width=0.4\columnwidth]{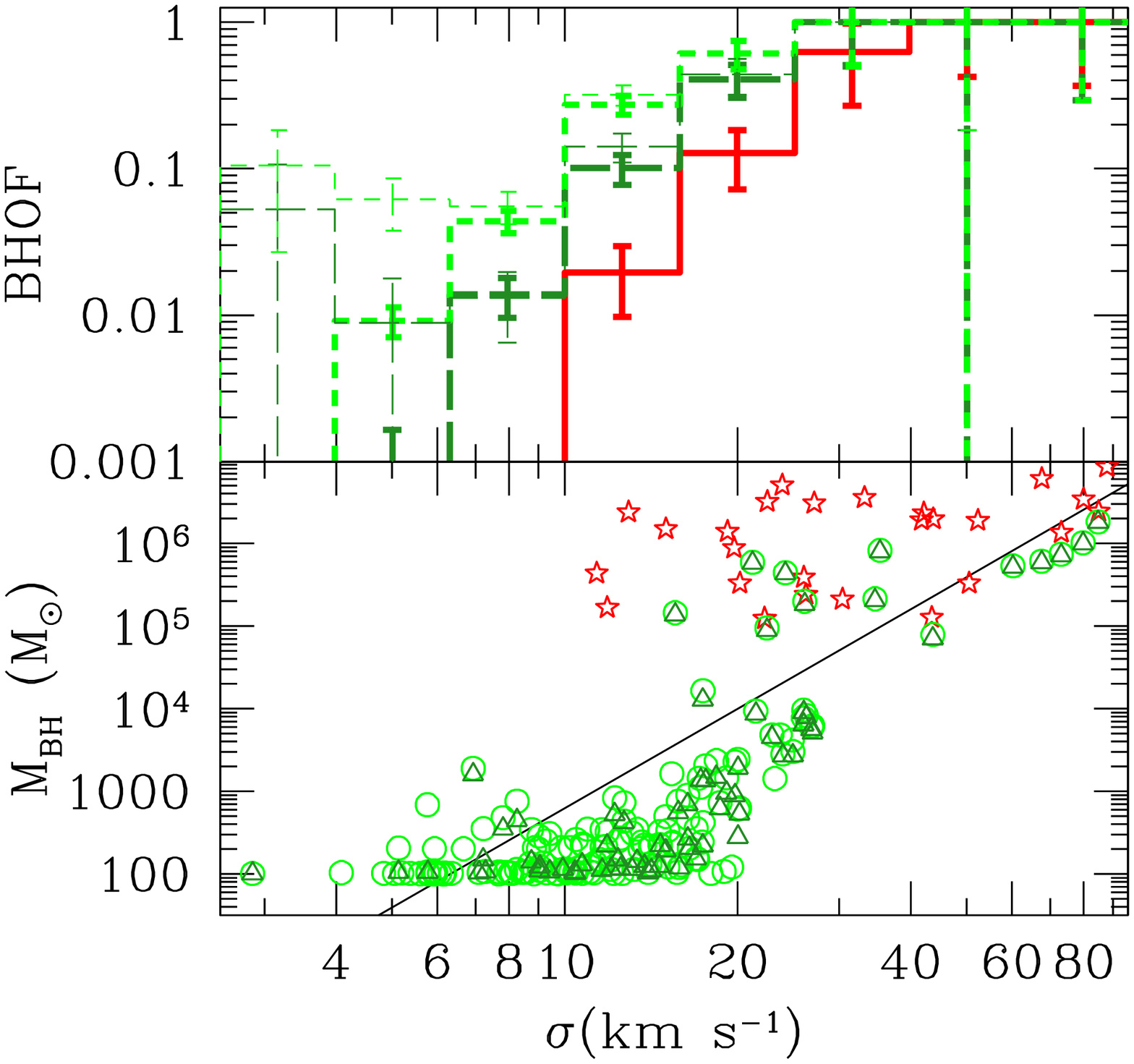}
\includegraphics[width=0.4\columnwidth]{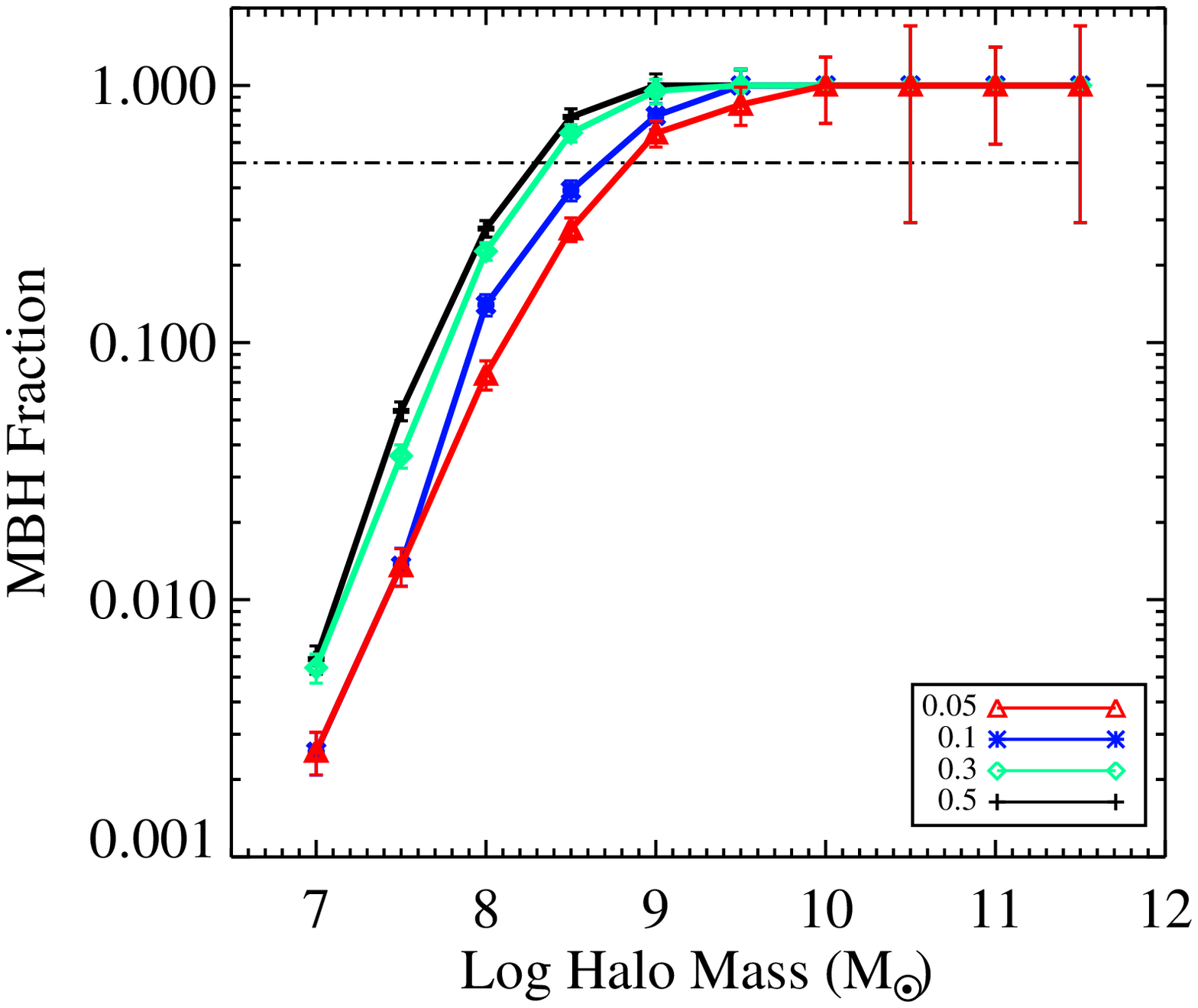}
\caption{Left: Occupation fractions and the correlation of MBH mass
and host velocity dispersion, $\sigma$, are different with different
MBH formation models.  Taken from \cite{VanWassenhove10}. Right: The
MBH-halo occupation fraction for a given halo mass at $z = 5$ in
cosmological simulations. Colored lines and symbols represent
simulations with different values of the efficiency of MBH formation
\cite{Bellovary11}. Reproduced by permission of the AAS
\label{fig:sandor}
}

\end{figure}

For example, \cite{VanWassenhove10} investigate the evolution of MBHs
in a Milky Way-like galaxy with two seed formation models -- one
resembling Pop III stars and one resembling direct collapse.  The main
difference between the models is effectively the mass of the seeds and
the formation efficiency, as massive seeds require larger halo masses
with specific angular momentum criteria compared to low-mass seeds.
The left panel of Figure \ref{fig:sandor} shows the occupation
fraction of MBHs (top), i.e., how many galaxies host an MBH, for
various Pop III models (green) and the direct collapse model
(red). This figure exemplifies where we may find a trace of the MBH
formation mechanism: at low galaxy mass.  While large systems always
host MBHs in either case, the Pop III model exhibits a larger
occupation fraction at smaller masses, due to the larger number of Pop
III black holes formed in the early universe with respect to the
direct collapse case.  An effect is seen in the MBH-host galaxy
relation as well; MBHs lurk in low mass galaxies but do not undergo
any significant growth over their lifetimes.  The mass of MBHs today
is still very close to the original seed mass (bottom). These objects
exist as intermediate mass black holes hiding in dwarf galaxies, and
are unlikely to be observed because of their extremely low predicted
accretion rates, and their tiny dynamical region of influence.

A critical question for MBH seed formation is therefore ``how many
dwarf galaxies host MBHs?''  While the answer currently appears to be
``very few,'' one must keep in mind that such objects would be
extremely difficult to observe.  Gas surface densities are low in
dwarfs, so accretion rates would be extremely small, even taking into
account gas lost from stars \cite{Volonteri2011}.  The surface density
of stars is also quite low, so detecting MBHs using stellar dynamics
is challenging given that there are so few stars in the radius of
influence \cite{VanWassenhove10}.  In some cases, such as the Large
Magellanic Cloud, the dynamical center is not well constrained, so it
is unclear where we should look to find an MBH.  Since dwarf galaxies
exhibit cored density profiles, the MBH may not even inhabit the exact
center of the galaxy, but may ``slosh'' around within the central core
region. Given these challenges, only a few dwarf galaxies have been
found to host AGN.  For example, the dwarf elliptical Pox 52
\cite{Barth04} and the dwarf irregular Henize 2-10 \cite{Reines11}
each host an MBH of around $10^5$ and $10^6$ \msun, respectively.

The occupation fraction of MBHs in galaxies is a key clue to the
mechanism of seed formation.  In cases where the formation efficiency
of MBH seeds is high, one expects to find seeds in a larger fraction
of galaxies, regardless of whether these seeds have grown into SMBHs.
Such an effect is indeed seen in cosmological simulations where seed
formation efficiency is varied \cite{Bellovary11} (Figure
\ref{fig:sandor}, right).  Observing the MBH halo occupation is
challenging, because for the most part we can only observe {\em
active} black holes in galaxies.  Still, measuring the active fraction
of MBHs in galaxies provides a lower limit on the MBH occupation
fraction.  Nuclear activity due to MBHs has been detected in 32\% of
the late-type galaxies in the Virgo Cluster, exclusively in galaxies
with mass $M_{halo} > 10^{10}$ \msun~ \cite{Decarli07}. For early-type
galaxies in Virgo, nuclear activity exists in 3-44\% of galaxies with
mass less than $10^{10}$ \msun, and 49 - 87\% of galaxies with mass
greater than $10^{10}$ \msun~ \cite{Gallo08}.  A similar study of
early-type field galaxies gives an occupation fraction of $45 \pm 7\%$
\cite{Miller12}.  Further studies will need to extend down to lower
mass galaxies as well as higher redshifts.  Data at $z \sim 1$ from
DEEP2 and AEGIS may help provide some constraints (see \cite{Yan11}).

\section{The Interplay Between Black Holes and Their Host Galaxies}

\begin{figure}[htb]
\includegraphics[width=0.4\columnwidth]{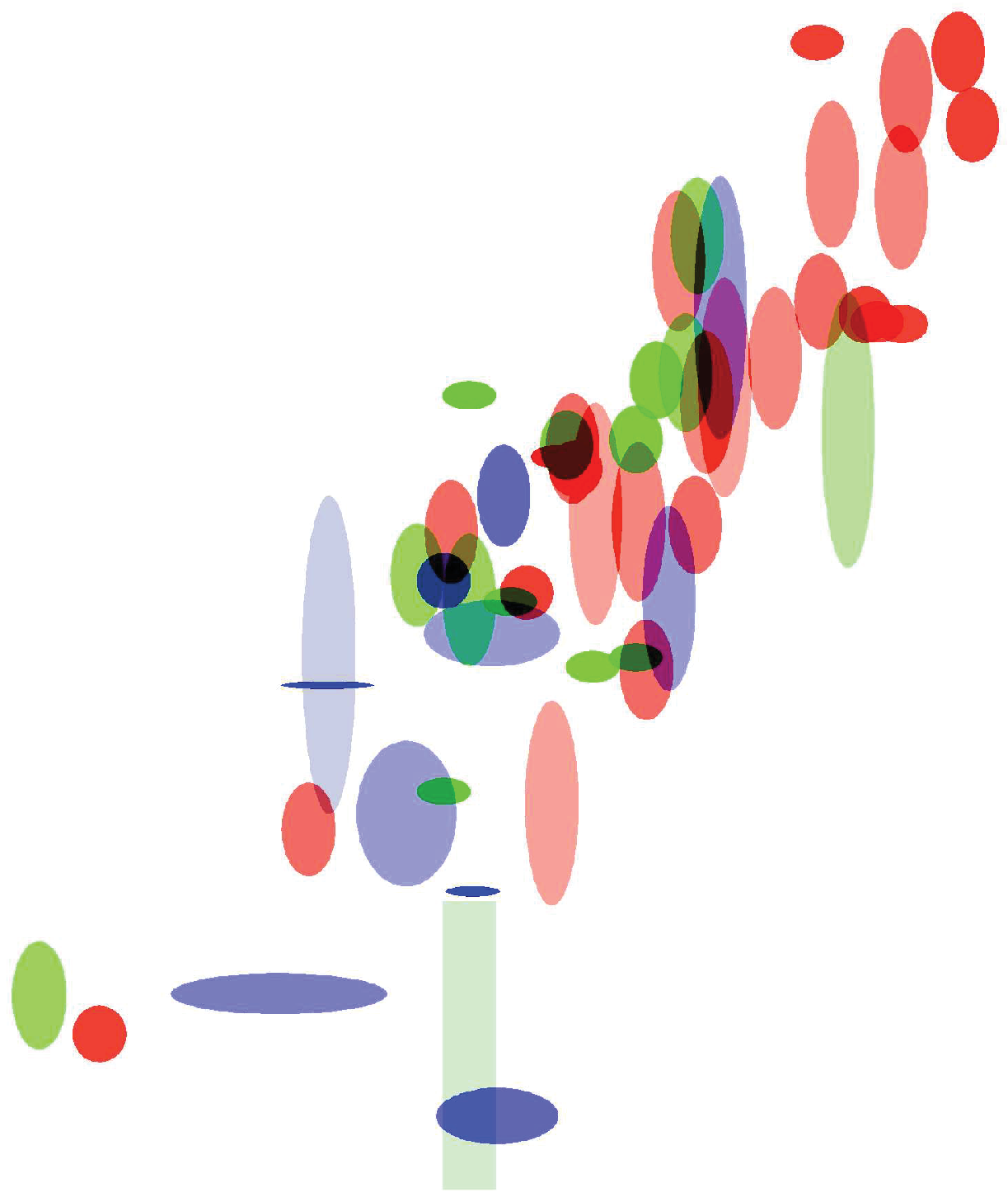}
\includegraphics[width=0.4\columnwidth]{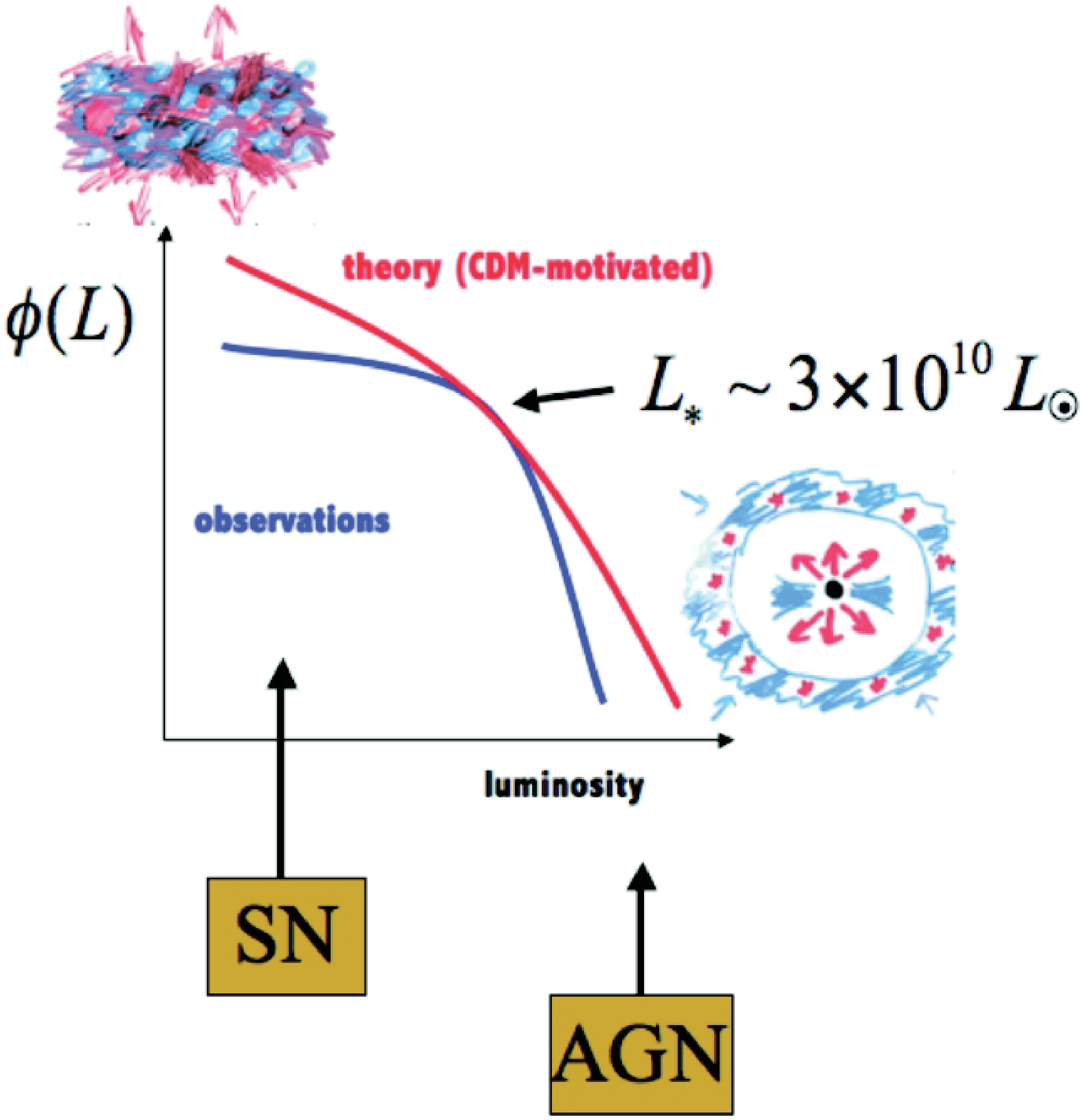}
\caption{Left: the masses of MBHs in nearby galaxies are found to
scale with many properties of the host galaxies, including the
velocity dispersion, as shown here.  Figure taken from
\cite{Gultekin09}.  Reproduced by permission of the AAS. Right:
cartoon depicting how we think that different types of feedback affect
the relationship between galaxies and dark matter halos. Supernovae
are believed to be responsible for limiting the formation of stars in
small galaxies, while AGN feedback is suggested to be shutting off
star formation in large galaxies. Figure taken from \cite{Silk11}.
\label{fig:gultekin}
}
\end{figure}

Today's MBH masses are found to scale with the properties of their
hosts, as described in the Introduction, thus determining MBH - host
scaling relations (e.g., Fig.~\ref{fig:gultekin}, left).  These
relations extend over several orders of magnitude in MBH mass, though
scatter does increase at both the low and high mass ends
\cite{Gultekin09,Greene10,McConnell11}.  It is also not entirely clear
what sets this relationship: is it a connection between the MBH and
the stellar properties of the bulge? Or, is the driving property the
global potential well of the galaxy, or even the host dark matter
halo?

%
%

Studying MBH scaling relations at higher redshifts may help
understanding how the relationships are established, but measurements
are fraught with difficulty.  Current studies yield mixed results,
with several authors finding that at high $z$ MBHs are more massive
for a given velocity dispersion
\cite{Peng06,Woo08,Bennert11,Targett12}, but others finding the
opposite \cite{Alexander08} or no evolution at all \cite{Sarria10}.
Part of the difficulty in deciphering the data is the difficulty of
obtaining reliable MBH mass estimates.  Robust methods that are used
in the local universe (i.e. dynamical measurements, reverberation
mapping) are not possible for more distant objects.  MBH masses must
be estimated using broad line widths; at higher redshifts, different
lines must be used for this measurement, and the correlation between
masses derived by H$\alpha$, C IV and Mg II has a large amount of
scatter \cite{Greene10}.  Bulge velocity dispersion measurements are
straightforward for nearby galaxies, but at higher redshifts a bulge
mass or luminosity estimate becomes much more difficult due to surface
brightness dimming, inclination effects, and dust extinction.  These
difficulties are compounded with the selection effects inherent in the
large surveys which are used for target selection.  If our surveys are
simply picking out objects which are the most luminous, and in turn
have the highest masses for a given host property, the MBH-host
relations will appear to evolve toward more massive MBHs at higher
redshift -- a type of Malmquist bias \cite{Lauer07b}.  These factors
compounded make observational determinations of the evolution of the
MBH scaling relations extremely challenging \cite{VS2011}.

While MBHs are harbored in the central regions of their galaxies,
evidence is mounting that their mere existence can affect a number of
larger-scale galaxy properties.  The primary culprit is postulated to
be feedback, or the effect of the energy released upon gas accretion
by the MBH \cite{Silk1998}.  Numerous studies have reported on the
repercussions of {\em stellar} feedback
\cite{Murray05,Oppenheimer08,Hopkins11} and it should not be
surprising that MBHs may act similarly.  In fact, AGN feedback has
been advocated as a solution for the underabundance of star forming
galaxies with respect to many predictions of theoretical models of
galaxy formation, making ellipticals ``red and
dead''\cite{Croton06,Schawinski2007}.  The discrepancy between the
simulated halo mass function and the observed galaxy luminosity
function is perhaps the best demonstrator of feedback effects.  In
Figure \ref{fig:gultekin} (right), we can see that the two curves
diverge at low and high masses.  The low-mass divergence is postulated
to be caused by stellar feedback; supernova winds drive baryons out of
small galaxies fairly efficiently, but as galaxies become larger the
winds do not have enough energy to escape the potential well.  At high
galaxy masses, AGN feedback may be strong enough to transport baryons
out of massive halos.  This loss of baryons decreases the star forming
potential of a galaxy, hence the divergence in the luminosity function
from the halo mass function.  Observationally, however, the situation
appears much more complex. In the host galaxies of moderate-luminosity
AGN, star formation typically precedes MBH accretion, suggesting that
in early phases supernova feedback may suppress MBH growth
\cite{Schawinski2009,Davies2007}, or that mass loss from
newly formed stars is what fuels MBHs \cite{Ciotti2007,Wild2010}.  Further
explorations of how MBHs are fueled, and how star formation and AGN
activity are linked on small scales \cite{Davies2004} or along cosmic
history \cite{Mullaney2012}, are needed before the complex interplay
of MBHs and their hosts is understood.

Another observed consequence of MBH feedback and its effect on a host
galaxy is AGN winds.  Such winds are observed on all scales, from very
close to the MBH in broad absorption line quasars, to galaxy-wide
distances (e.g. \cite{Tremonti07}).  These outflows cause the transport of gas from the galaxy center to its halo, 
or possibly out of the system entirely.  For example, \cite{Nesvadba08} observed
three quasars with radio jets at $z \sim 2-3$ and found outflow
velocities of 800-1000 km s$^{-1}$ and ionized gas masses on the order
of $10^{10}$ \msun.  Molecular AGN-powered outflows have been detected
in more local galaxies by \cite{Feruglio10,Alatalo11}, who find that
the mass outflow rate is enough to deplete the molecular gas reservoir
in about a dynamical time.  Clearly such a rapid gas mass loss will
drastically affect the evolution of a galaxy.  Determining how much
gas is lost and how quickly is a complex question of AGN luminosity,
gas properties, and the mechanism of the energy coupling to the gas.
These questions are non-trivial and cannot be generalized; a full
range of parameter space must be explored in simulations and
observations in order to make headway here.

\section{Modeling Black Holes and Galaxies in a Cosmological Context}

In order to capture the evolutionary process of MBHs from their birth
to the present day, one must embed the physics of MBHs in a
cosmological setting.  Cosmological simulations are key for tracing
MBH and galaxy populations throughout their lifetimes, gaining
meaningful statistics, and exploring the main drivers of their
evolution.

\subsection{Semi-Analytical  Models}
Analytical models of MBH formation and evolution have been developed
as described in sections 2 and 3. These models have been embedded in
so-called semi-analytical models of galaxy formation. Semi-analytical
codes study the evolution of the baryonic content of galaxies (gas,
stars, black holes) upon a skeleton given by the dark matter halo
merger history, which can be extracted from N-body cosmological
simulations, or built through Monte Carlo techniques, given initial
conditions (cosmological parameters, power spectrum of density
fluctuations).  Semi-analytical techniques are based on physically
motivated recipes, which describe the physical processes occurring
during the evolution of galactic structures.  For instance, one can
compute several properties of the MBH population, given by the
combination of the birth rate, death rate (MBH-MBH mergers), and
accretion rate on each MBH. The latter provides a duty cycle for the
active MBHs (quasars and AGNs).  The code requires that we specify
physical properties for the evolving galaxies and MBHs, e.g., density
profiles of halos and galaxies, accretion rate onto MBHs (e.g.,
Eddington, Bondi), accretion disc properties (e.g., Shakura--Sunyaev),
and what the triggers for accretion and its termination are (e.g.,
galaxy mergers, feedback on the host). We also need to specify the
physical laws governing the evolution of MBH binaries (e.g., dynamical
friction, stellar dynamical equations).  As an example
\cite{Volonteri03,Volonteri08,Volonteri10} models are based on
comprehensive set of quantities of theoretical interest, outlined in
the scheme below.

{  \footnotesize
\vspace{0.1truecm}
\begin{center}
\begin{tabular}{|l|l|l}
Input: ~~~~~~~~~~~~{\Large $\rightarrow$} & Modelization: ~~~~{\Large $\rightarrow$} & Observables:\\
- Cosmology 	     	 & - MBH birth rate						 & - AGN LF at $z\ge0$\\
- Galaxy properties: density          	 &- MBH merger rate						 &- AGN redshift distributions\\
profiles, dynamical friction, stellar 	 & - MBH ejection rate                      & - MBHs mass functions \\
dynamical equations  					 & - Growth rate 			 & - Occupation fraction of MBHs \\
- Accretion properties: Eddington ratio,  & - Duty cycle			 & - LISA event rates, masses, z\\
Spectral energy distribution & & - X-ray background\\

\end{tabular}

\vspace{0.1truecm}
  \end{center}
}

The theoretical properties can be cast in terms of {\em{observables}}
to be tested against current and future data. For each model we can
calculate Luminosity Functions (LFs), number counts and redshift
distributions of AGN at different redshifts and wavelengths,
contribution to cosmic backgrounds, and event rates for gravitational
wave detectors. The input parameters of semi-analytical models are
very similar to those of hydrodynamical simulations, as described in
section 4.2.

Semi-analytical models are very efficient and provide an excellent
insight to physical mechanisms, allowing to explore a large parameter
space in short computational times. The main weakness of
semi-analytical models is the lack of spatial information (except for
the code PINOCCHIO, that uses Lagrangian techniques to derive the
spatial relationship between merging halos as well \cite{Pinocchio}),
the necessity of sticking to a rigid scheme that applies to all
systems in the sample, and the inability of modeling systems that do
not have smooth, analytical properties (e.g., galaxy mergers). The
main advantages of semi-analytical models is that they are very
efficient and provide an excellent insight to physical mechanisms,
allowing to explore a large parameter space in short computational
times. Because of the extremely high computational cost, in fact,
cosmological simulations do not allow for extended parameter space
exploration or large volume sampling at high spatial resolution
\cite{springel2005,DiMatteo08}.

\subsection{Numerical Simulations}

The ability to self-consistently model halo growth, gas accretion,
star formation, feedback processes, hydrodynamics, and MBH activity
makes these models the most sophisticated theoretical tool we have to
explore galaxy and MBH evolution.  Of course, there are also
limitations to this method; the need to model substantial volumes with
a large dynamic range limits the spatial and mass resolution, and
specific models of how, e.g., stars form and explode as supernovae, or
MBH form and accrete must be implemented as ``sub-grid" physics, as
they cannot be explicitly resolved. To give a sense of the problem,
the size of the disc of the Milky Way is tens of kpc, and our galaxy
is embedded in a dark matter halo of size $\sim$ Mpc.  The scale of
the region where an MBH dominates gravity is $\sim$ tens of pc. The
event horizon (Schwarzschild radius) of an MBH is of order $\mu$pc
scale. Therefore if one wanted to resolve the plunge of particles into
an MBH in a galaxy like the Milky Way would need a dynamical range of
12 orders of magnitude.  On scales of the event horizon, additionally,
simulations would require full general relativity. Finally, the
difference in timescales between galaxies and MBHs translates into
enormously different required timesteps. While on galactic scales
timescales are of order of at least $10^4-10^5$ yrs, close to the
horizon of a BH timescales are of order of days or hours, as material
moves close to the speed of light. This said, one attempts to capture
as well as possible the most important physical mechanisms to
understand the formation and evolution of MBHs in a cosmological
setting.

\subsubsection{Seed Formation.}

The first generation of cosmological smoothed particle hydrodynamics
(SPH) simulations placed one MBH in the center of each massive galaxy
by running an on-the-fly halo finder and identifying halos with a
predetermined threshold mass (usually $10^9$ or $10^{10}$ \msun) and
which did not already host an MBH
\cite{Sijacki06,Sijacki07,DiMatteo08,Okamoto08,Booth09,McCarthy10}.
MBHs of a set initial mass (generally $\sim 1 - 5 \times 10^5$ \msun)
are then fixed to the center of the halo, and are allowed to grow
through gas accretion and mergers with other MBHs.  Some properties of
the MBH population are correctly captured by this method, in the sense
that they do not depend on the initial mass and formation efficiency
of MBHs. For instance the luminosity function of quasars, which
measures the properties of the most massive MBHs in the most massive
galaxies at any given time, is not very sensitive to the initial
properties of the population; these properties may affect only the
faint end, which is not well determined observationally. Therefore
this strategy works well for understanding the global properties of
the evolved population of MBHs at late times, however, this approach
is unlikely to capture the physics of MBH formation.

\begin{figure}[htb]
\includegraphics[width=0.4\columnwidth]{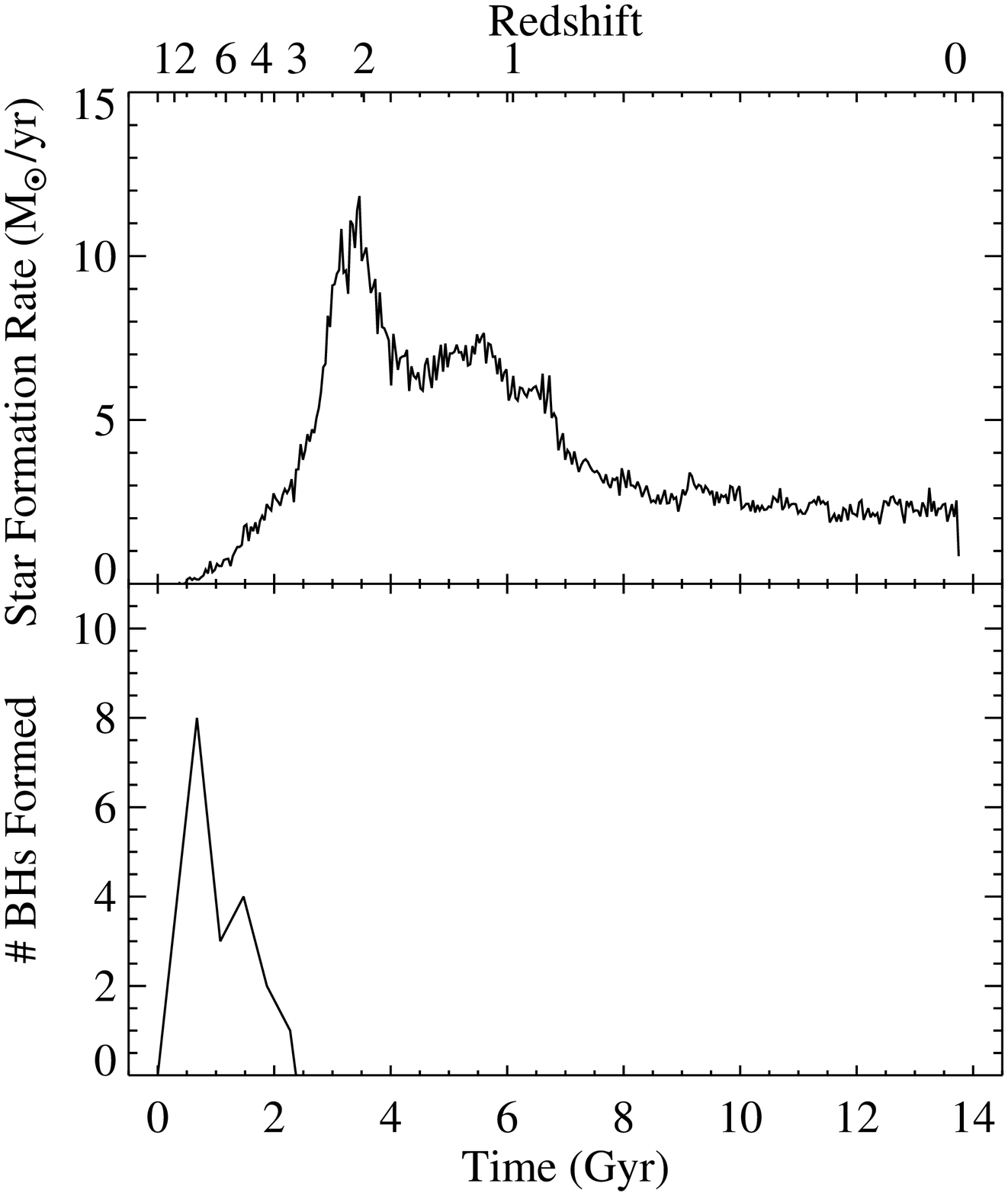}
\includegraphics[width=0.5\columnwidth]{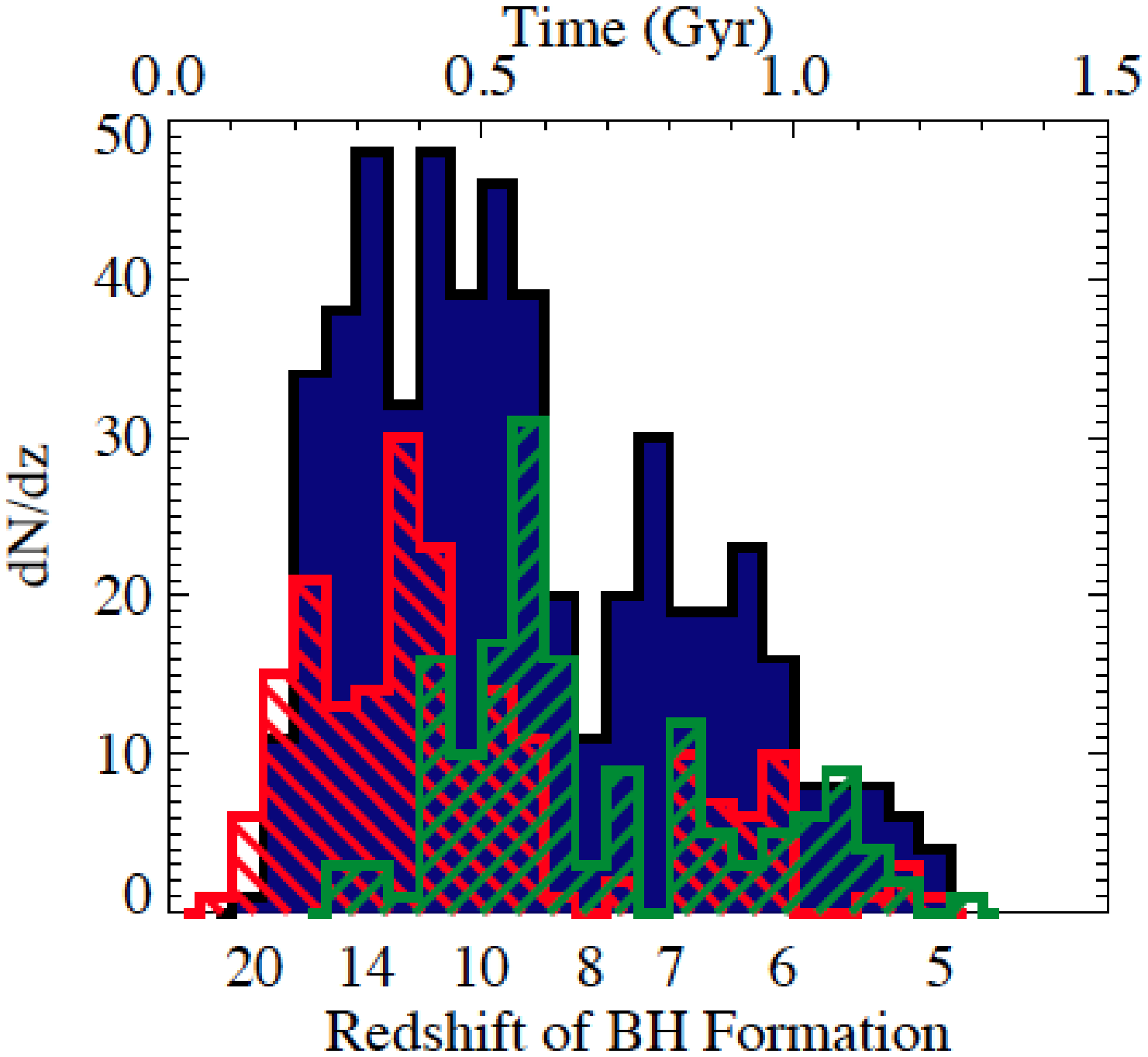}
\caption{Left: MBH and star formation rates from one of our Milky
Way-size galaxy. MBH formation is truncated due to contamination by
heavy elements, while stars continue forming. Right: Time of formation
of MBHs in a dwarf galaxy (green), a Milky-Way galaxy (red) and a
massive elliptical (blue). We can see the trend of earlier onset of
MBH formation in the more biased cosmic overdensities. The rate of
seed formation dwindles around z = 5 because supernova explosions have
enriched the Universe with heavy elements.  Both panels adapted from
Bellovary (2011). \label{seeds} }

\end{figure}

An alternate approach is taken by \cite{Bellovary11}, who employ a
zero-metallicity requirement as well as density and temperature
thresholds for MBH formation.  If a particle meets density,
temperature, and metallicity criteria, it is given a probability of
forming an MBH seed.  This probability mimics the efficiency of MBH
seed formation and can be adjusted freely.  As a result of this
method, more than one MBH can exist per galaxy, though the most
massive ones tend to form early on and remain at the galaxy centers,
near the regions of early dense star formation. Fig.~\ref{seeds}
exemplifies how this MBH formation model naturally forms MBHs only at
high redshift. MBH formation ends when the Universe becomes enriched
with heavy elements created by the first generations of stars. Heavy
elements have more efficient cooling via line emission, which in turn
favors formation of individual stars over the efficient collection of
gas conducive to MBH formation.

Adaptive mesh refinement (AMR) simulations use a different approach.
In \cite{Dubois12} MBHs are spawned from maximally refined regions
which are Jeans unstable.  These regions must exceed a set gas density
threshold as well as a stellar fraction threshold, to ensure that MBHs
form from dense gas after the formation of the first stars.  To
maintain only one MBH per galaxy, MBHs are not allowed to form within
some $r_{min}$ of another MBH.


\subsubsection{Accretion.}

The physical processes which govern the accretion of gas by MBHs take
place on scales with size of order the solar system, roughly 12 order
of magnitude smaller than the size of a whole galaxy.  It is not
surprising then that large-scale simulations of galaxy formation must
make assumptions about how gas travels from larger, resolvable scales
to the MBH.  Phenomena such as angular momentum transfer, magnetic
fields, and general relativity must be ignored.  Cosmological
simulations by \cite{Levine10} demonstrated that mass inflow toward an
MBH is extremely chaotic on small scales, rendering a broad
characterization of large-scale MBH accretion estimates extremely
difficult.  Still, with some clever sub-grid modeling we can hope to
create a basic representation of reality.

The most common approach to modeling MBH accretion is a method based
on the Bondi-Hoyle formalism:

\begin{equation}
\dot{M} = \frac{4\pi \alpha G^2 M_{BH}^2 \rho^2}{(c_s^2 + v^2)^{3/2}}
\end{equation}

where $\rho$ and $c_s$ are the local gas density and sound speed, $v$
is the relative speed between the MBH and the surrounding gas, and
$\alpha$ is a dimensionless parameter.  The value of $\alpha$ in the
precise Bondi-Hoyle formula ranges between 0.25 - 1.12, depending on
the equation of state of the gas.  However, simulators have taken this
parameter to be extremely flexible, ranging from one \cite{DiMatteo08}
to 1000 \cite{Sijacki07} to a varying function of density
\cite{Booth09}.  The reason for this variation is due to the inability
to resolve the Bondi radius (and thus the high gas densities one
expects at this radius).  If the gas density at the Bondi radius is
underestimated, the MBH's growth will be artificially stunted.
Various authors have adjusted the parameter $\alpha$ such that their
cosmological simulations reproduce the local $M - \sigma$ relation and
the black hole density evolution over cosmic time.

One can argue that using Bondi-Hoyle accretion as an estimate for the
true accretion rate is grossly wrong.  Efficient accretion occurs in
discs, not spherical systems.  While Bondi-Hoyle may be accurate in
the case of hot, low density, spherically symmetric gas such as in
elliptical galaxies, this case is not a common one and is certainly
not the situation when MBHs are accreting most of their mass at high
redshift.  Indeed, recent simulations show that Bondi-Hoyle accretion
is interrupted if a moderate radiation feedback component is
introduced \cite{Milosavljevic09,Barai12} (though see \cite{Park11} for a solution to this problem involving episodic accretion).  While the use of this
model may seem questionable, it can also be argued that it is
reasonable for the scales which can be resolved in an SPH cosmological
simulation.  At a distance of a few 100 parsecs from an MBH, any gas
which is nearby can be assumed to fall in toward the MBH within a
reasonable amount of time.  Measuring the gas properties at a large
scale can give us a basic estimate of what the accretable gas near the
MBH might be like.  While we will surely benefit from more
sophisticated models in the future, current Bondi-Hoyle models do
broadly reflect reality and give us a basic idea of how MBHs interact
with the gas in galaxy centers.


More experimentation with MBH accretion has been done in
non-cosmological simulations.  \cite{DeBuhr10} developed a viscous
disk model of MBH accretion which, when used in idealized simulations
of galaxy mergers, reproduces properties of the local $M - \sigma$
relation.  Alternatively, \cite{Power10} created an ``accretion disk
particle'' which mimics an MBH and its accretion disk.  When low
angular momentum gas particles come within the disk radius, they are
captured and accreted within an accretion timescale.  \cite{Dotti2007}
use a simulation of a circumnuclear disc with a resolution of 0.1 pc
to study the properties of the gas that gets bound to an MBH. They
define weakly bound, bound, and strongly bound particles according to
the relationship between the total (sum of the kinetic, internal and
gravitational) energy per unit mass, $E$, and the gravitational
potential, $W$, of the MBH (e.g., strongly bound particles have $E<0.5W$,
weakly bound particles have $E<0.25W$, bound particles have
$E<0$). They find that dynamical effects, e.g., the tidal field of
the MBH, affect the distribution of particles over the duration of
their simulations.  None of these models have been implemented in a
cosmological simulation; interesting future work would be to see how
they fare in different galaxy environments, and how the accretion
estimates compare to those derived from Bondi-Hoyle.

\subsubsection{Feedback}

Modeling feedback is critical for simulations of MBH evolution, but as
in the case of accretion, the actual physical processes involved are
completely unresolved.  The most common sub-grid model for MBH
feedback is an isotropic thermal energy deposition \cite{DiMatteo05}.
If one assumes that a fraction $\epsilon_r$ of the mass accreted by
the MBH is converted into energy (usually $\epsilon_r = 10\%$), and a
fraction $\epsilon_f$ of that energy couples to the surrounding gas,
the feedback energy deposition rate is equal to

\begin{equation}
\dot{E} = \epsilon_r \epsilon_f \dot{M} c^2
\end{equation}

This method of feedback modeling has been used in the majority of
cosmological simulations, with a value of $\epsilon_f = 0.01- 0.15$
\cite{Sijacki07,DiMatteo08,Booth09,Bellovary10}.  \cite{Sijacki07}
employ a second feedback mechanism as well, to represent mechanical
feedback on cluster scales with hot buoyant bubbles.  Such simulations
have had success in reproducing the observed MBH - host scaling
relations as well as the black hole space density evolution.

%

 More advanced feedback models are possible in extremely high
resolution simulations, but to achieve this one must sacrifice
resolution for simulation size.  For example, \cite{Jeon11} have run a
1 Mpc$^3$ cosmological simulation to $z \sim 10$ which includes
radiative feedback effects from Population III stars and MBHs.  They
are able to include X-ray photoionization heating processes as a
result of accretion onto the MBH, and conclude that the growth of MBH
seeds from Population III stars is severely hampered by these feedback
effects.  Some more recent works with AMR simulations have also
incorporated more sophisticated forms of feedback models.
\cite{Dubois12} include the thermal feedback model mentioned earlier
in this section, but also invoke mechanical feedback at times when
accretion rates drop below 1\% of the Eddington rate (to represent a
radio jet).  With this model, mass, momentum and energy are
distributed within a cylinder oriented along the angular momentum axis
of the nearby particles.

A model developed by \cite{Kim11} utilizes a radiative mode and a
mechanical mode of feedback.  For the radiative mode, the high
resolution of the simulation allows for a full three-dimensional
radiative transfer calculation to be done, with the MBH as the
emitting source.  Photons are emitted with a temperature of 2 keV and
are then able to ionize, heat, and exert momentum onto the surrounding
gas.  For the mechanical mode, mass is periodically injected along a
jet axis with a velocity of 6000 km/s.  This model is the most highly
sophisticated to date and is extremely promising; however a fully
self-consistent cosmological simulation including MBH seed formation
at high redshift has not yet been done.

\begin{center}
\begin{figure}[htb]
\includegraphics[width=6in]{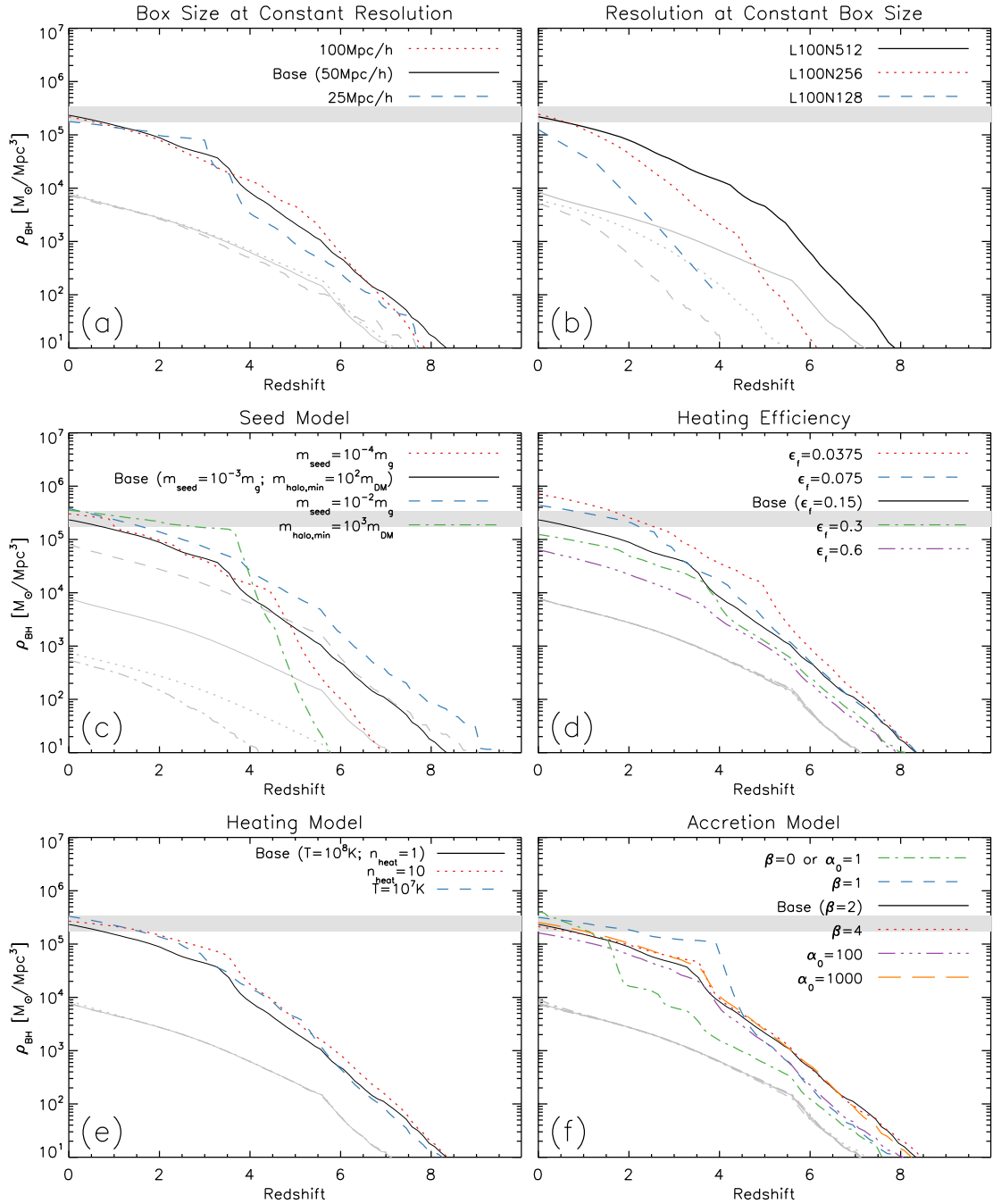}
\caption{An example parameter study of MBHs in cosmological
simulations. Shown here is MBH mass density vs. redshift for a wide
range of physical models and parameters.  The shaded grey line
represents the $z = 0$ value.  \cite{Booth09}
\label{fig:booth}
}

\end{figure}
\end{center}

\subsubsection{Results}

Cosmological simulations do an excellent job of broadly representing
the growth and evolution of MBHs and their host galaxies.  Simulated
galaxies have realistic star formation histories, and their predicted
colors are what we expect to see in observed populations.  We can
reproduce the MBH - host scaling relations and the black hole space
density at $z = 0$.  However, many quantities which are not directly
observable are also difficult to quantify in simulations.  For
example, the MBH mass density is constrained locally \cite{Shankar04},
but at higher redshifts there are no constraints at all.  Varying
simulation parameters such as resolution, box size, feedback
efficiency, or accretion models can give different answers at high
redshift while agreeing at $z = 0$.  This is illustrated in Figure
\ref{fig:booth}, which shows the result of a large parameter space
exploration and how the MBH density is affected by varying one at a
time.  Clearly there are a wide range of ``acceptable'' models that
reproduce the observed $z = 0$ value but vary greatly at higher
redshifts.

All in all, cosmological simulations have given us many clues to how
MBHs form, grow, and evolve with their host galaxies.  The overall
picture is fairly clear; MBH seeds form early on in small halos, grow
through mergers and gas accretion, and affect their surroundings via
feedback.  However, the finer details of the physical processes
involved are still beyond our reach.  Higher resolution simulations
and more sophisticated modeling are required in order to answer our
remaining questions.

\section{Outstanding Questions and Future Developments}

There has been much progress to report on the formation and evolution
of MBHs, and much more progress will happen in the near future. While
many theoretical and observational advances have improved our
understanding of how MBHs form and evolve in galaxies, many questions
still remain. Theorists have brought forward several models of MBH
formation (section 2) but it is very hard to find a direct smoking gun
that can pinpoint the right mechanism, or even rule them all out. From
the observational point of view we have to rely on secondary
indicators (e.g., the occupation fraction of MBHs, their masses in
dwarf galaxies, constraints from growth time). From the theoretical
point of view, we are either limited to analytical models, or to
numerical simulations that can barely resolve the scales where these
processes operate, in a well known trade-off between the resolution
one can reach and the size of the galaxy one can model.

The co-evolution between galaxies and MBHs is a very active topic of
research, and many complementary approaches are likely to provide
important clues in the near future. Properties of AGN hosts from local
galaxies up to $z\sim 2$ (e.g., SDSS, zCOSMOS) are being coupled with
estimates of MBH masses (through line widths \cite{Vestergaard06}),
with more and more attention to possible biases \cite{Kelly10}. Clever
techniques are being used to study not only the most luminous quasars
but also less luminous sources \cite{Treister11,Mullaney2012}. Ongoing
work using integral-field spectroscopy
\cite{Davies2010,Storchi2010,Muller-Sanchez2011,Riffel2011} on nearby
Seyfert galaxies is revealing spatially resolved distributions and
kinematics of the gas that feeds the MBH and fuels star formation. The
James Webb Space Telescope and Euclid, successors of the Hubble Space
Telescope, and the Atacama Large Millimeter Array will zoom in on the
highest quasars, and may eventually bring us to the edge of the Dark
Ages when the first MBHs and galaxies formed.

\ack MV acknowledges funding support from NASA, through award ATP
NNX10AC84G; from SAO, through award TM1-12007X, from NSF, through
award AST 1107675, and from a Marie Curie Career Integration grant,
PCIG10-GA-2011-303609.


\end{document}